\documentclass[aps,prl,showpacs,twocolumn]{revtex4}
\usepackage{epsfig,amsmath,color,amssymb}
\usepackage[10pt]{moresize}

\begin{document}

\title{Synchronized Independent Narrow-band Single Photons
and Efficient Generation of Photonic Entanglement}

\author{Zhen-Sheng Yuan$^{1,2}$}
\author{Yu-Ao Chen$^1$}
\author{Shuai Chen$^1$}
\author{Bo Zhao$^1$}
\author{Markus Koch$^1$}
\author{Thorsten Strassel$^1$}
\author{Yong Zhao$^1$}
\author{Gan-Jun Zhu$^1$}
\author{J\"{o}rg Schmiedmayer$^{1,3}$}
\author{Jian-Wei Pan$^{1,2}$}
\address{$^1$Physikalisches Institut, Ruprecht-Karls-Universit\"{a}t Heidelberg, Philosophenweg 12,
69120 Heidelberg, Germany}
\address{$^2$Hefei National Laboratory for Physical Sciences at Microscale and Department of Modern Physics,
University of Science and Technology of China, Hefei, Anhui 230026, China}
\address{$^3$Atominstitut der \"{o}sterreichischen Universit\"{a}ten, TU-Wien, A-1020 Vienna Austria}
\date{\today}

\begin{abstract}

We create independent, synchronized single-photon sources with built-in quantum memory
based on two remote cold atomic ensembles. The synchronized single photons are used to
demonstrate efficient generation of entanglement.  The resulting entangled photon pairs
violate a Bell's Inequality by 5 standard deviations. Our synchronized single photons
with their long coherence time of 25 ns and the efficient creation of entanglement serve
as an ideal building block for scalable linear optical quantum information processing.

\end{abstract}



\maketitle

Synchronized generation of either deterministic and storable single photons or entangled photon
pairs is essential for scalable linear optical quantum information processing (LOQIP). With the
help of quantum memory and feed-forward, one can thus achieve long-distance quantum communication
\cite{BriegelPRL1998, PanNature2001, DuanNature2001} and efficient quantum computation
\cite{KnillNature2001, RaussendorfPRL2001, NielsenPRL2004, BrownePRL2005}. Very recently,
interfering synchronized independent single photons \cite{KaltenbaekPRL2006} and entangled photon
pairs \cite{YangPRL2006} have been experimentally achieved with two pulsed spontaneous parametric
down-conversion sources pumped by two synchronized but mutually incoherent femto-second lasers.
However, due to the absence of quantum memory for broad-band (a few nm) single photons no feedback
was applied in the above experiments, single photons or entangled photon pairs were merely
generated probabilistically in each experimental run, i.e. with a small probability $p$. Thus, in
an experiment concerning manipulation of $N$ synchronized single (or entangled) photon sources, the
experimental efficiency will decrease exponentially with the number of sources (proportional to
$p^{N}$). Moreover, the short coherence time of down-converted photons ($\sim$ a few hundred fs,
defined by the bandwidth of interference filters) also makes hard the overlap of photon wavepackets
coming from two distant sites. These two drawbacks together make the above experiments
inappropriate for scalable LOQIP.

Following a recent proposal for long-distance quantum communication with atomic ensembles
\cite{DuanNature2001} (see also the improved schemes \cite{ChenzbQuantPh2006}), it is
possible to generate narrow-band single photons or entangled photon pairs in a
deterministic and storable fashion. In the past years, significant experimental
progresses have been achieved in demonstration of quantum storage and single-photon
sources \cite{ChaneliereNature2005,EisamanNature2005}, and even entanglement in number
basis for two atomic ensembles has been demonstrated experimentally
\cite{ChouNature2005}. Moreover, deterministic narrow-band single-photon sources have
been demonstrated most recently with the help of quantum memory and electronic feedback
circuits \cite{ MatsukevichPRL2006deterSP, RiedmattenPRL2006, ChenShuaiPRL2006}.

In this Letter, we develop further the techniques used in Ref. \cite{ChenShuaiPRL2006} to implement
synchronized generation of two independent single-photon sources from two remote atomic ensembles
loaded by magneto-optical traps (MOT). The two synchronized single photons are further used to
demonstrate efficient generation of entangled photon pairs. Since our single-photon sources are
generated in-principle in a deterministic and storable fashion, with the help of feed-forward the
experimental methods can be used for scalable generation of photonic entanglement. Moreover,
compared to the short coherence time of down-converted photons in Refs. \cite{KaltenbaekPRL2006,
YangPRL2006} the coherence time of our synchronized narrow-band single photons is about 25 ns, four
orders longer, which makes it much easier to overlap independent photon wavepackets from distant
sites for further applications of LOQIP. Finally, it is worth noting that the read and write lasers
used for different single-photon sources are fully independent to each other. The synchronization
was achieved by separate electronic signals generated by the control electronics.

\begin{figure}[htp]
\includegraphics[width=8.0cm]{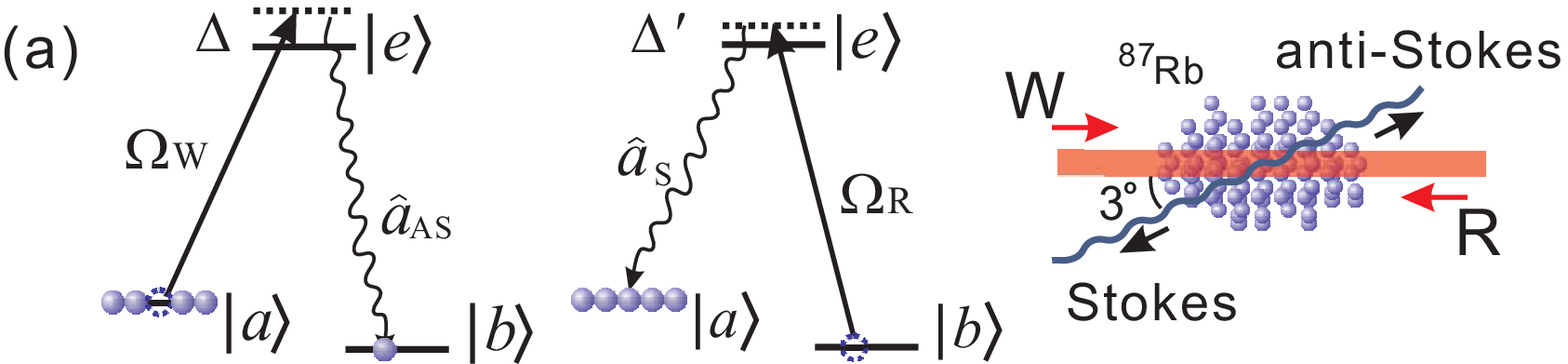}\\
\vspace{2mm}
\includegraphics[width=8.0cm]{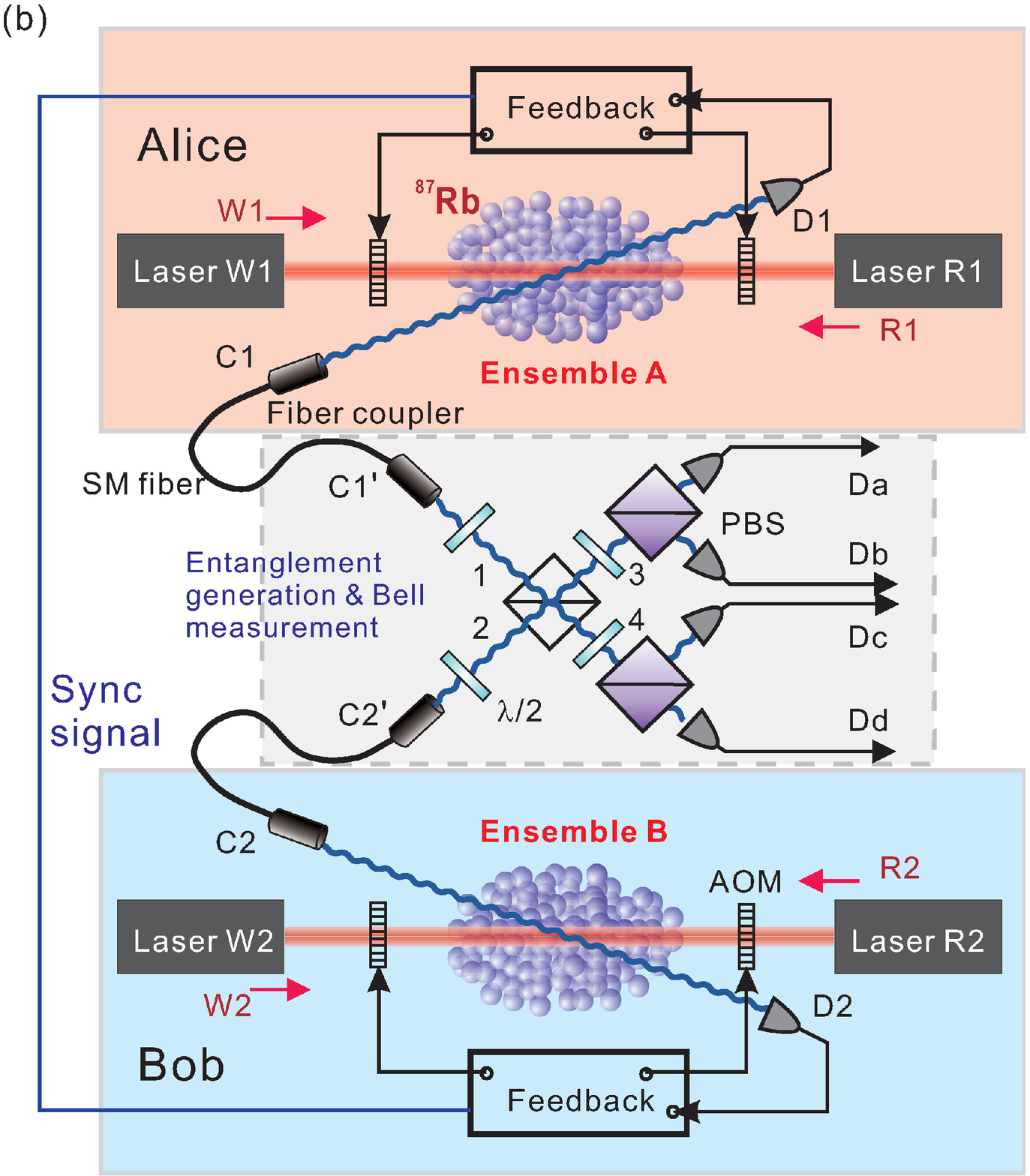}
\caption{Illustration of the relevant energy levels of the atoms and arrangement of laser
beams (a) and the experimental setup (b). (a) $^{87}$Rb atoms are prepared in the initial
state $|a\rangle$. A write pulse $\Omega_{\mbox{\tiny W}}$ with the detuning of
$\Delta=10$ MHz and a beam diameter about 400 $\mu$m is applied to generate the spin
excitation and an accompanying photon of the anti-Stokes field $\hat{a}_{\mbox{\tiny
AS}}$ with a beam diameter about 100 $\mu$m. The mode $\hat{a}_{\mbox{\tiny AS}}$, tilted
$3^\circ$ from the direction of the write beam, is coupled in a single-mode fiber (SMF)
and guided to a single-photon detector. Waiting for a duration $\delta t_{\mbox{\tiny
R}}$, a read pulse is applied with orthogonal polarization and spatially mode-matched
with the write beam from the opposite direction. The spin excitation in the atomic
ensemble will be retrieved into a single photon of the Stokes field $\hat{a}_{\mbox{\tiny
S}}$, which propagates to the opposite direction of the field $\hat{a}_{\mbox{\tiny AS}}$
and is also coupled in SMF. (b) Alice and Bob each keeps a single-photon source at two
remote locations. As elucidated in Ref. \cite{ChenShuaiPRL2006}, Alice applies write
pulses continuously until an anti-Stokes photon is registered by detector D1. Then she
stops the write pulse, holds the spin excitations and meanwhile sends a synchronization
signal to Bob and waits for his response (This is realized by the feedback circuit and
the acousto-optic modulators, AOM). In parallel Bob prepares a single excitation in the
same way as Alice. After they both agree that each has a spin excitation, each of them
will apply a read pulse simultaneously to retrieve the spin excitation into a light field
$\hat{a}_{\mbox{\tiny S}}$. The two Stokes photons propagate to the place for
entanglement generation and Bell measurement. They overlap at a 50:50 beam splitter (BS)
and then will be analyzed by latter half-wave plates ($\lambda$/2), polarized beam
splitters (PBS) and single photon detectors Da, Db, Dc, and Dd. \label{fig:setup}}
\end{figure}

The basic concept of our experiment is illustrated in Fig. \ref{fig:setup}. Atomic
ensembles collected by two MOT's 0.6 m apart function as the media for quantum memories
and deterministic single-photon sources. Each ensemble consists of about $10^8$ $^{87}$Rb
atoms. The two hyperfine ground states $|5S_{1/2},F=2\rangle$=$|a\rangle$ and
$|5S_{1/2},F=1\rangle$=$|b\rangle$ and the excited state
$|5P_{1/2},F=2\rangle$=$|e\rangle$ form a $\Lambda$-type system
$|a\rangle$-$|e\rangle$-$|b\rangle$. The atoms are initially optically pumped to state
$|a\rangle$. Shining a weak classical \textit{write} pulse with the Rabi frequency
$\Omega_{\mbox{\ssmall W}}$ into the atoms, creates a superposed state of the anti-Stokes
field $\hat{a}_{\mbox{\ssmall AS}}$ and a collective spin state of the atoms,
\begin{eqnarray}\label{eqn:state}
|\Psi\rangle\sim|0_{\mbox{\ssmall AS}}0_b\rangle+ \sqrt{\chi}|1_{\mbox{\ssmall AS}}1_b\rangle+
\chi|2_{\mbox{\ssmall AS}}2_b\rangle+O(\chi^{3/2}),
\end{eqnarray}
where $\chi\ll 1$ is the excitation probability of one spin flip, and $|i_{\mbox{\ssmall
AS}}i_b\rangle$ denotes the $i$-fold excitation of the anti-Stokes field and the
collective spin. Ideally, conditioned on detecting one and only one anti-Stokes photon, a
single spin excitation is generated in the atomic ensemble with certainty. In practice,
considering photon loss in the detection, this condition can be fulfilled by keeping
$\chi\ll 1$ so as to make the multi excitations negligibly small. After a controllable
time delay $\delta t_{\mbox{\ssmall R}}$ (in the order of the lifetime $\tau_c$ of the
spin excitation), another classical \textit{read} pulse with the Rabi frequency
$\Omega_{\mbox{\ssmall R}}$ is applied to retrieve the spin excitation and generate a
photon in the Stokes field $\hat{a}_{\mbox{\ssmall S}}$. If the retrieve efficiency
reaches unity, the Stokes photon i  s no longer probabilistic because of the quantum
memory and feedback control \cite{ChenShuaiPRL2006, RiedmattenPRL2006,
MatsukevichPRL2006deterSP}, which now can serve as a  deterministic single-photon source.
As shwon in Fig. \ref{fig:setup}, Alice and Bob both have such a source. They prepare
collective spin excitations independently and the one who finishes the preparation first
will wait for the other while keeping the collective spin excitation in her/his quantum
memory. After they agree that both have finished the preparation, they retrieve the
excitations simultaneously at anytime they want within the lifetime of the collective
state. Therefore the retrieved photons arrive at the beam splitter with the required
timing.

Compared to a probabilistic photon source, the present implementation with atomic ensembles
contributes a considerable enhancement to the coincidence rate of single photons coming from Alice
and Bob. For instance, we consider a similar setup but without feedback circuit, where Alice and
Bob apply write and read in every experimental trial and thereafter measure the four-fold
coincidence of anti-Stokes and Stokes photons in the four channels D1, D2, C1 and C2. Assume the
probability to have an anti-Stokes photon in channel D1 (D2) is $p_{\mbox{\ssmall A\!\!\:S}1}$
($p_{\mbox{\ssmall A\!\!\:S}2}$) and the corresponding retrieve efficiency for conversion of the
spin excitation to a Stokes photon coupled into channel C1 (C2) is $\gamma_1(\delta
t_{\mbox{\ssmall R}})$ [$\gamma_2(\delta t_{\mbox{\ssmall R}})$], then the probability of four-fold
coincidence is $p_{4c}=p_{\mbox{\ssmall A\!\!\:S}1} \gamma_1(\delta t_{\mbox{\ssmall R}})
p_{\mbox{\ssmall A\!\!\:S}2}
 \gamma_2(\delta t_{\mbox{\ssmall R}})$. This has to be compared with using the feedback circuits shown in Fig.
\ref{fig:setup}, where we can apply at most $N$ (limited by the lifetime of the quantum memory and
the speed of the feedback circuit) write pulses in each trial. Then the probability of four-fold
coincidence becomes
\begin{eqnarray}\label{eqn:enhance}
P_{4c}&=\Big\{ \sum_{i=0}^{N-1}p_{\mbox{\ssmall A\!\!\:S}1}(1-p_{\mbox{\ssmall
A\!\!\:S}1})^i \sum_{j=i}^{N-1}p_{\mbox{\ssmall A\!\!\:S}2}(1-p_{\mbox{\ssmall A\!\!\:S}2})^j  \nonumber\\
& \times \gamma_2(\delta t_{\mbox{\ssmall R}}) \gamma_1[(j-i)\cdot\delta t_{\mbox{\ssmall
W}}+\delta t_{\mbox{\ssmall R}}]\Big\}+\Big\{\cdots\Big\}_{1\leftrightarrow2},
\end{eqnarray}
where $\delta t_{\mbox{\ssmall W}}$ is the time interval between the sequential write pulses
\cite{ChenShuaiPRL2006} and $\{\cdots\}_{1\leftrightarrow2}$ is the same as the first term with
index 1 and 2 being exchanged. Assume $p_{\mbox{\ssmall A\!\!\:S}1}\ll 1$ and $p_{\mbox{\ssmall
A\!\!\:S}2}\ll 1$ and a long lifetime $\tau_c$, we obtain $P_{4c}\sim N^2 p_{\mbox{\ssmall
A\!\!\:S}1} \gamma_1(\delta t_{\mbox{\ssmall R}}) p_{\mbox{\ssmall A\!\!\:S}2}  \gamma_2(\delta
t_{\mbox{\ssmall R}})$ for a definite number $N$. So the probability of four-fold coincidence is
enhanced by $N^2$ for each trial. For our case $p_{\mbox{\ssmall A\!\!\:S}1}\approx
p_{\mbox{\ssmall A\!\!\:S}2}=2.0\times10^{-3}$ (the relevant cross correlation $g_{\mbox{\ssmall
A\!\!\:S,S}}^{(2)}=30$), $N=12$, $\tau_c\sim 12~\mu$s, $\delta t_{\mbox{\ssmall W}}=800$ ns,
$\delta t_{\mbox{\ssmall R}}=400$ ns, and $\gamma_1(0)\approx\gamma_2(0)=8\%$, the enhancement is
136.

The four lasers in Fig. \ref{fig:setup} are independently frequency stabilized. The
linewidths of W1 and R1 are about 1 MHz while those of W2 and R2 are about 5 MHz of the
full width at half maximum (FWHM). However, they will be broadened to more than 20 MHz
because the laser pulse modulated by the AOM is a Gaussian-like profile with width about
40 ns FWHM. The linewidth of the retrieved single photons is determined mainly by the
linewidth and intensity of the read lasers. So we try to make the profile of the two
independent read pulses identical to each other.

In order to verify that the two Stokes photons coming from Alice and Bob are
indistinguishable, we let them overlap at a BS with the same polarization (horizontal in
our case) and measure the quantum interference indicated by the the Hong-Ou-Mandel (HOM)
dip \cite{HongPRL1987}. Having observed the high visibility of HOM dip in both time
domain and frequency domain, we are confirmed that the two independent photons are
indistinguishable. Then we put one of the two photons to vertical polarized before they
enter the BS. By coincidence measurement at the two outputs of the BS, we generate the
Bell state
$|\Psi^{-}\rangle_{12}=\frac{1}{\sqrt{2}}(|H\rangle_1|V\rangle_2-|V\rangle_1|H\rangle_2)$,
which is verified by the measurement of violation of Bell's Inequality.

\begin{figure}
\includegraphics[width=4.0cm]{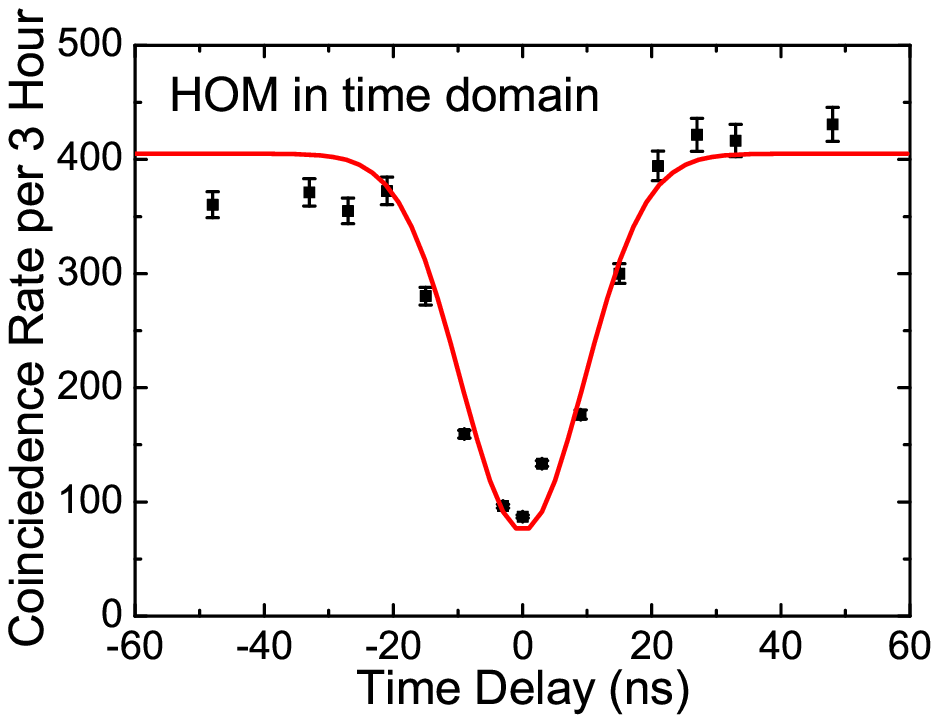}
\includegraphics[width=4.0cm]{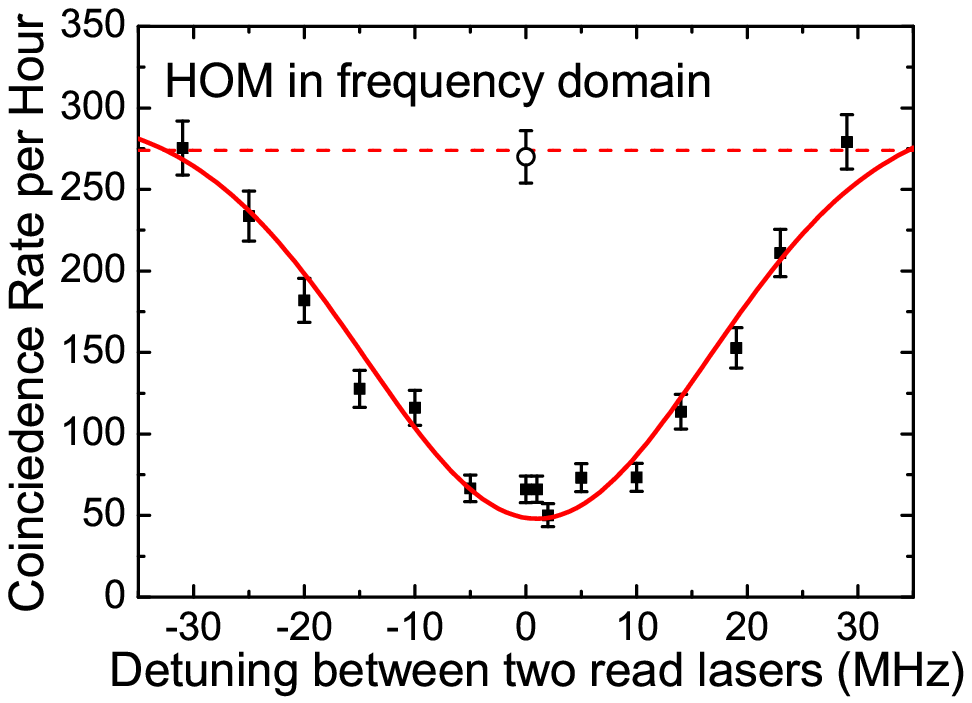}
\caption{Hong-Ou-Mandel dips in time domain (left panel) and frequency domain (right
panel). The circle in the right panel was obtained by setting the polarization of the two
photons perpendicular to each other and zero detuning between two read lasers. The
Gaussian curves that roughly connect the data points are only shown to guide the eye. The
dashed line shows the plateau of the dip. Error bars represent statistical errors, which
are $\pm$ one standard deviation.\label{fig:hom}}
\end{figure}

\textit{The measurement of HOM dip}. We did two measurements to obtain the HOM dip in time domain
and frequency domain respectively. To make the photons indistinguishable, the polarizations of the
anti-Stokes photons were set to horizontal with two half-wave plates before they enter the BS as
shown in Fig. \ref{fig:setup}. The other two half-wave plates after the BS were set to $0^\circ$.

In the first measurement, we measured the four-fold coincidence among detectors D1, D2, Da and Dd
while changing the time delay between the two read pulses (Fig. \ref{fig:hom}, left panel). The
excitation probabilities $p_{\mbox{\ssmall A\!\!\:S}1}\approx p_{\mbox{\ssmall
A\!\!\:S}2}=2.0\times10^{-3}$. The coincidence rate is varied with the delay. Ideally, there should
be completely destructive interference if the wavepackets of the two photons overlap perfectly.
However, it is hard to make the two wavepackets absolutely identical or exactly overlapped in
practice. We obtained the visibility of the dip $V=(C_{\mbox{\ssmall plat}}-C_{\mbox{\ssmall
dip}})/C_{\mbox{\ssmall plat}}=(80\pm1)\%$, where $C_{\mbox{\ssmall plat}}$ is the non-correlated
coincidence rate at the plateau and $C_{\mbox{\ssmall dip}}$ is the interfering coincidence rate at
the dip. The asymmetry of the profile at negative delay and positive delay shows that the two
wavepackets are (a) not perfectly identical, (b) not symmetric themselves. Assume the HOM dip is a
Gaussian-type profile, we estimate the coherence time is 25$\pm$1 ns FWHM.

In the second measurement, we measured the four-fold coincidence among detectors D1, D2, Da and Dd
while changing the frequency detuning between the two read pulses (Fig. \ref{fig:hom}, right
panel). It is the first time to measure HOM dip in frequency domain at single-photon level. The
excitation probabilities are $p_{\mbox{\ssmall A\!\!\:S}1}\approx p_{\mbox{\ssmall
A\!\!\:S}2}=3.0\times10^{-3}$, higher than those in time domain. Because of the limit of the
current setup, the detuning can be varied from $-30$ MHz to 30 MHz. In order to verify the
coincidence rate at largest detuning reached the plateau of HOM dip, we measured the coincidence by
setting the polarization of the two photons perpendicular to each other and zero detuning between
the two read lasers (shown as a circle in Fig. \ref{fig:hom}). The consistence of this data with
those two at largest detunings shows that we have achieved the plateau of HOM dip. The visibility
is $(82\pm3)\%$ which agrees well with that obtained in time domain. The width of the HOM dip is
35$\pm$3 MHz FWHM, in accordance with the coherence time 25 ns. Therefore, the narrow-band
characteristic of the present source is verified directly by the HOM dip in the frequency domain.

Besides the imperfect overlap of the single-photon wavepackets, the two-photon components
in each of the single-photon sources affect the visibility as well. The quality of
single-photon source is characterized by the anti-correlation parameter
$\alpha=2P_{\mbox{\ssmall I\!\!\:I}}/P_{\mbox{\ssmall I}}^2$ \cite{ChenShuaiPRL2006},
where $P_{\mbox{\ssmall I}}$ ($P_{\mbox{\ssmall I\!\!\:I}}$) is the probability of
generating one (two) photon(s) for each source (the higher orders are negligible small).
If the two wavepackets do not overlap at all, there is no interference between them. Then
we obtain the non-correlated coincidence rate $C_{\mbox{\ssmall plat}}=P_{\mbox{\ssmall
I}}^2/2+P_{\mbox{\ssmall I\!\!\:I}}$ between Da and Dd. If they overlap perfectly, there
is destructive interference leading to a coincidence rate $C_{\mbox{\ssmall
dip}}=P_{\mbox{\ssmall I\!\!\:I}}$. So the visibility of the HOM dip is $V=1/(1+\alpha)$.
In our experiment, $\alpha=0.12$ for the source prepared later (the spin excitation is
retrieved immediately) and $\alpha=0.17$ for the source prepared earlier (it has to wait
for the other one). This leads to an average visibility of 87\%. In the frequency domain,
the average visibility is around 83\% because of higher excitation probabilities.

\textit{Efficient entanglement generation}. As shown in Fig. \ref{fig:setup}, we set orthogonal
polarizations (horizontal and vertical) of the Stokes photons with the two half-wave plates before
the BS. Then the state of the two photons will be projected to $|\Psi^{-}\rangle_{12}$ if there is
coincidence between the two output port 3 and 4. With another two half-wave plates and two PBS
after the BS, the entanglement of the two photons can be verified by a Clauser-Horne-Shimony-Holt
(CHSH) type inequality \cite{ClauserPRL1969}, where $S\leq 2$ for any local realistic theory with
\begin{eqnarray}\label{eqn:chsh}
S=|E(\theta_1,\theta_2)-E(\theta_1,\theta_2^\prime)-E(\theta_1^\prime,\theta_2)-
E(\theta_1^\prime,\theta_2^\prime)|.
\end{eqnarray}
Here $E(\theta_1,\theta_2)$ is the correlation function where $\theta_1$ and
$\theta_1^\prime$ ($\theta_2$ and $\theta_2^\prime$) are the measured polarization angles
of the Stokes photon at port 3 (4). The observed values of the correlation functions are
listed in Table \ref{tab:bell} resulting in $S=2.37\pm0.07$, which violates Bell's
Inequality by 5 standard deviations. This clearly confirms the quantum nature of the
entanglement state.

With our imperfect sources we do not create a perfect $|\Psi^{-}\rangle_{12}$. If we consider the
two photon component in the photon sources the created state will be:
\begin{eqnarray}\label{eqn:effstate}
|\Psi_{\mbox{\ssmall eff}}\rangle_{12}=\left\{
\begin{array}{l}
P_{\mbox{\ssmall I}}^2/2,~1/\sqrt{2}(|H\rangle_1|V\rangle_2-|V\rangle_1|H\rangle_2); \\
P_{\mbox{\ssmall I\!\!\:I}}/2,~|H\rangle_1|H\rangle_2; \\
P_{\mbox{\ssmall I\!\!\:I}}/2,~|V\rangle_1|V\rangle_2.
\end{array}
\right.
\end{eqnarray}
From the quality of the single photons generated from the two ensembles, $\alpha=0.12$, $0.17$ and
Eq. (\ref{eqn:effstate}), we estimate the expected violation of Bell's Inequality is around $2.3$,
which is in good agreement with our measured value. It is interesting to note that a violation of
Bell's Inequality needs a single photon source with $\alpha<0.24$ according to
Eq.(\ref{eqn:effstate}). In order to minimize $\alpha$, further improvements, e.g., a higher
optical couple efficiency, a lower photon loss, a lower excitation probability and a higher
retrieve efficiency, will be made in our future investigations.


\begin{table}\caption{Correlation functions $E$ and the resulting $S$. \label{tab:bell}}
\begin{tabular}{ccc|c}
\hline
$E$  & $\theta_1=0^\circ$ & $\theta_1^\prime=45^\circ$ & $S$ \\
\hline
$\theta_2=22.5^\circ$ & $-0.613\pm 0.037$ & $0.575\pm0.039$ & \\
$\theta_2^\prime=-22.5^\circ$ & $0.606\pm 0.038$ & $0.579\pm0.039$ & $2.37\pm 0.07$\\
\hline
\end{tabular}
\end{table}
In conclusion, we realized synchronized generation of narrow-band single photons with two remote
atomic ensembles. The Hong-Ou-Mandel dip was observed in both time domain and frequency domain with
a high visibility for independent photons coming from two distant sites, which shows the
indistinguishability of these photons. By virtue of quantum memories and feedback circuits, the
efficiency of generating entangled photon pairs was enhanced by a factor of 136, which claims our
single-photon source as a promising candidate for the future implementation of scalable quantum
computation based on linear optics \cite{KnillNature2001, RaussendorfPRL2001, NielsenPRL2004,
BrownePRL2005}. The present spatially-distributed independent single-photon sources (with fully
independent write and read lasers) are pre-requirements for the long-distance quantum communication
\cite{BriegelPRL1998, ChenzbQuantPh2006}. The narrow-band property (which makes the overlap of the
photon wavepackets at the order of nanoseconds) of single photons and high efficiency of
entanglement generation also profit the present source to serve as an ideal candidate for large
scale communications, e.g., satellite-based quantum communication. There is still potential to
improve our single-photon source. We can improve the retrieve efficiency close to unity by
increasing the optical density of the atomic ensemble. A better compensation of the stray magnetic
field will help the extension of the lifetime up to 100 $\mu$s. If we want an even longer lifetime,
a good solution is to confine the atoms in an optical trap, which also benefits to a much higher
optical density.

This work was supported by the Deutsche Forschungsgemeinschaft (DFG), the Alexander von Humboldt
Foundation, the Deutsche Telekom Stiftung, the Konrad-Adenauer Stiftung, the LGFG and the CAS.

\textit{Note added}.-- During the final phases of our experiment we became aware of two
related experiments \cite{FelintoNPhysics2006,ChanelierePRL2007}.

\end{document}